\documentclass[3p,times,twocolumn]{elsarticle}
\usepackage{ifthen}
\usepackage{xspace}
\usepackage{booktabs}
\usepackage{multirow}
\usepackage{ctable}
\usepackage{calc}
\usepackage{relsize}
\usepackage[colorlinks=false,pdfborder={0 0 0}]{hyperref}

\newcommand{\format}{1}
\newcommand{\ifdraft}[2]{\ifthenelse{\equal{\format}{0}}{#1}{#2}}
\newcommand{\ifarxiv}[2]{\ifthenelse{\equal{\format}{1}}{#1}{#2}}
\newcommand{\ifnproc}[2]{\ifthenelse{\equal{\format}{2}}{#1}{#2}}

\ifarxiv{
  \usepackage{amsmath}
  \usepackage{amssymb}
  \journal{Nuclear Physics B Proceedings}
}{
  \usepackage{ecrc}
  \usepackage{amsmath}
  \usepackage{amssymb}
  \journalname{Nuclear Physics B Proceedings Supplement}
  \jnltitlelogo{Nuclear Physics B Proceedings Supplement}
  \jid{nuphbp}
  \volume{00}
  \firstpage{1}
  \runauth{P. Ilten}
}

\ifdraft{
  \usepackage[pdf,clean]{svg}
  \usepackage{lineno}
  \linenumbers
}{}

\newcommand{\tls}{\ensuremath{\tau}-leptons\xspace}
\newcommand{\tl}{\ensuremath{\tau}-lepton\xspace}
\newcommand{\tauto}{\ensuremath{\tau^- \rightarrow \nu_\tau}}
\newcommand{\tn}{\ensuremath{{\nu_\tau}}\xspace}
\newcommand{\mnb}{\ensuremath{{\bar{\nu}_\mu}}\xspace}
\newcommand{\enb}{\ensuremath{{\bar{\nu}_e}}\xspace}
\newcommand{\ha}{\ensuremath{H}\xspace}
\newcommand{\hb}{\ensuremath{h^0}\xspace}
\newcommand{\hc}{\ensuremath{H^0}\xspace}
\newcommand{\hd}{\ensuremath{A^0}\xspace}
\newcommand{\he}{\ensuremath{H^\pm}\xspace}
\newcommand{\z}{\ensuremath{Z}\xspace}
\newcommand{\dy}{\ensuremath{\gamma^*/Z}\xspace}
\newcommand{\w}{\ensuremath{W}\xspace}
\newcommand{\rh}{\ensuremath{\rho}\xspace}
\newcommand{\de}{\ensuremath{\mathcal{D}}\xspace}
\newcommand{\me}{\ensuremath{\mathcal{M}}\xspace}
\newcommand{\cp}{\ensuremath{\mathcal{CP}}\xspace}

\newcommand{\pythia}{{\sc Pythia}\xspace}
\newcommand{\herwig}{{\sc Herwig++}\xspace}
\newcommand{\tauola}{{\sc Tauola}\xspace}

\begin{document}

\begin{frontmatter}
  
  \title{Tau Decays in \pythia $8$}
  \author{Philip Ilten}
  \address{School of Physics, University College Dublin}
  
  \begin{abstract}
    As of version $8.150$ of \pythia, the isotropic decay model of
    \tls has been replaced with sophisticated \tl decay machinery. The
    decays and spin correlations for \tls in \pythia $8$ are
    described, including the spin correlation algorithm, the available
    \tl production processes, the \tl decay models, the user
    interface, and the implementation.
  \end{abstract}
  

\end{frontmatter}

\section{Introduction}

The role of \tls in Higgs boson measurements~\cite{sm} and beyond the
Standard Model searches~\cite{bsm} is becoming increasingly important,
due to the enhanced coupling of the \tl in many of these physics
models. Consequently, it is necessary for current Monte Carlo
generators to ensure accurate modeling of \tl decays.

Prior to version $8.150$ of \pythia~\cite{pythia}, \tl decays in
\pythia were performed using a leptonic or generic hadronic current
matrix element without including spin correlations, and more
sophisticated \tl modeling was only possible through external packages
such as \tauola~\cite{tauola}. Now, in \pythia version $8.150$ and
above, fully modeled hadronic currents with spin correlations are
available, based on prior \tl modeling work in \tauola and
\herwig~\cite{herwig}. Currently, all known \tl decays with a
branching fraction greater than $0.04\%$ are modeled.

In this review, the spin correlation algorithm used in \pythia for
\tls is described, as well as the available \tl production mechanisms,
summarized in Table~\ref{tab:production}, and \tl decays, given in
Table~\ref{tab:decay}. An introduction to the user interface, as well
as the technical implementation is given.

\section{Correlations}

The spin correlation algorithm used in \pythia for \tl decays is based
on the algorithm proposed by Collins~\cite{collins} and
Knowles~\cite{knowles}, and expanded by
Richardson~\cite{richardson}. The algorithm separates spin
correlations from the hard process, parton shower, and hadronization
phases of the Monte Carlo generator, while maintaining full
correlations, and can be broken into the following steps.

\begin{enumerate}
\item The $2 \to n$ hard process is generated according to its matrix
  element \me.\label{ite:hard}
\item One of the outgoing particles is selected and its helicity
  density matrix is calculated, 
  \begin{equation}
    \begin{aligned}
      \rh_{\lambda_j \lambda_j'}^j = ~&
      \rh_{\kappa_1\kappa_1'}^1\rh_{\kappa_2\kappa_2'}^2
      \me_{\kappa_1\kappa_2; \lambda_1 \ldots \lambda_n}
      \me_{\kappa_1'\kappa_2'; \lambda_1' \ldots \lambda_n'}^* \\
      & \times \prod_{k \neq j} \de_{\lambda_k\lambda_k'}^k \\
    \end{aligned}
  \end{equation}
  and the trace is normalized. Here, $\rh^{1,2}$ are the helicity
  density matrices of the incoming particles with helicity
  $\kappa_{1,2}$, \me is the matrix element with outgoing particle
  helicities $\lambda_k$, and $\de^k$ are the decay matrices of the
  outgoing particles, initialized to the identity. If the particle is
  from a decay, the helicity density matrix is the same as above but
  without $\rh^2$ and $\kappa_2$.
\label{ite:rho}
\item The selected particle is decayed using the weight,
  \begin{equation}
    \mathcal{W} = 
    \rh_{\lambda_0\lambda_0'}
    \me_{\lambda_0;\lambda_1 \ldots \lambda_n}
    \me_{\lambda_0';\lambda_1' \ldots \lambda_n'}^*
    \prod_{k = 1,n} \de_{\lambda_k\lambda_k'}^k
  \end{equation}
  where $\rho$ is the helicity density matrix of the decaying particle
  with helicity $\lambda_0$ and the decay matrix element \me.\label{ite:decay}
\item Steps~\ref{ite:rho} through~\ref{ite:decay} are performed until
  a decay is reached with no unstable particles.
\item The decay matrix of the last decayed particle is calculated,
  \begin{equation}
    \de_{\lambda_0\lambda_0'} = 
    \me_{\lambda_0;\lambda_1 \ldots \lambda_n}
    \me_{\lambda_0';\lambda_1' \ldots \lambda_n'}^* \prod_{k
      = 1,n} \de_{\lambda_k\lambda_k'}^k
  \end{equation}
  and the trace is normalized.\label{ite:de}
\item An undecayed particle from the decay above is randomly selected
  and steps~\ref{ite:rho} through~\ref{ite:de} are repeated.\label{ite:rec}
\item Step~\ref{ite:rec} is repeated until all unstable particles are
  decayed.
\end{enumerate}

\section{Production}

Spin correlated \tl decays can be generated in \pythia from a variety
of hard processes, summarized in Table~\ref{tab:production}. The spin
of the \tl can also be provided from an externally generated hard
process, or explicitly set for all \tls or \tls from a specified
parent.

For electroweak processes, spin correlations are fully handled if the
incoming fermions producing the $\gamma$, \z, or \w are known. If the
\w is not produced from fermions or the fermions are unavailable, the
\tl is given a polarization of $\mathcal{P} = -1$, e.g. $H \to \w \to
\tau\tn$. For $\gamma$ or \z hard processes not produced from fermions
or with unavailable fermion information, the $\gamma$ or \z is assumed
to be unpolarized, e.g. $H \to \z\z \to 4\tau$.

\begin{figure}
  \begin{center}
    \ifdraft
    {\includesvg[width=\columnwidth,name=fig1,pretex=\footnotesize]{production}}
    {\includegraphics[width=\columnwidth]{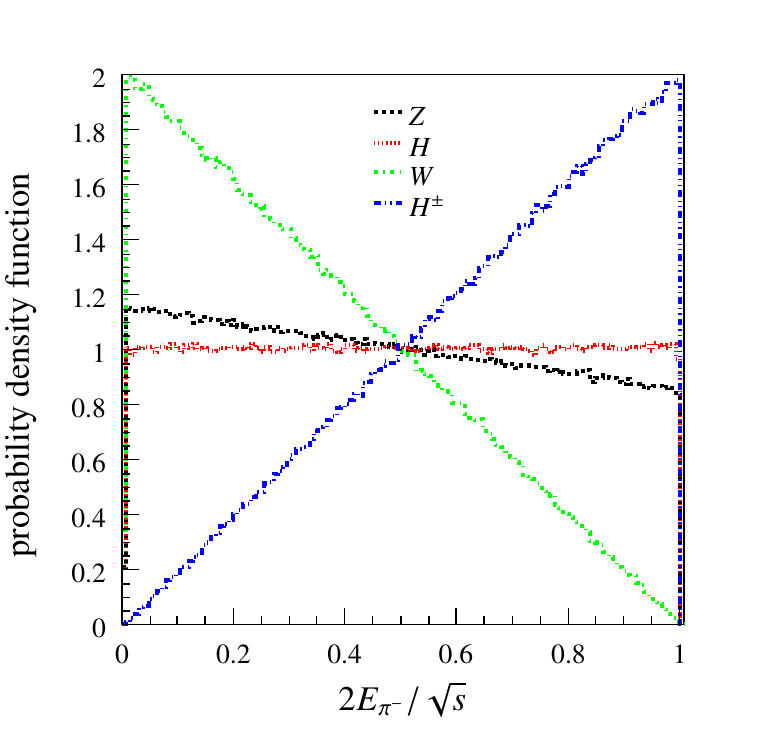}}
    \caption{Distribution of the fractional energy of the pion from a
      $\tauto \pi^-$ decay in the rest frame of the producing
      boson. The \tls are produced from \z (dashed black), \ha (dotted
      red), \w (dash-dotted green), and \he (dash-dot-dotted blue)
      bosons.\label{fig:production}}
  \end{center}
\end{figure}

Because the Higgs is spin zero, the production information is not
necessary for calculating \tl spin correlations. However, the type of
Higgs must be known. Currently, spin correlations of \tls produced
from the \cp-even (\ha, \hb, \hc), \cp-odd (\hd), or charged (\he)
Higgses are calculated.

The \tl can also be produced from a variety of $B$ and $D$ meson
decays. For these decays, the constituent quarks of the meson are
approximated, and the spin correlations are calculated using an
$f\bar{f}' \to W \to f\bar{f}'$ matrix element. For any \tl from an
unknown process, the \tl is assumed to be unpolarized.

The slope of the energy distribution for pions from $\tauto \pi^-$
decays in the rest frame of the parent bosons is proportional to the
average \tl polarization. In Figure~\ref{fig:production} this
distribution, from \pythia simulation, is given for \tls produced from
electroweak and Higgs processes. Note the the opposite polarization of
\tls produced from a \w and \he, the unpolarized \tls produced from
neutral Higgses, and the slight polarization of \tls produced from \z
bosons at a $pp$ collider.

\begin{table}[!h]
  \small
  \begin{center}
    \begin{tabular}{c|l}
      \toprule
      Type & 
      \multicolumn{1}{c}{Processes} \\
      \midrule[\heavyrulewidth]
      \multirow{3}{*}{electroweak} & $f\bar{f} \to \gamma \to f\bar{f}$, $f\bar{f} \to \z \to f\bar{f}$, \\
                                   & $f\bar{f} \to \dy \to f\bar{f}$, $f\bar{f}' \to \w \to f\bar{f}'$, \\
                                   & $\z \to f\bar{f}$, $\w \to f\bar{f}'$ \\
      \midrule
      \multirow{2}{*}{Higgs}       & $\ha \to f\bar{f}$, $\hb \to f\bar{f}$, $\hc \to f\bar{f}$\\
                                   & $\hd \to f\bar{f}$, $\he \to f\bar{f}'$ \\
      \midrule
      other                        & $B/D \to f\bar{f}' + X$ \\
      \bottomrule
    \end{tabular}
    \caption{Production mechanisms in \pythia for which full spin
      correlations of \tl decays are automatically
      calculated.\label{tab:production}}
  \end{center}
\end{table}

\section{Decays}

The matrix element for the decay of the \tl can be written as,
\begin{equation}
  \me = \frac{g_w^2}{8 m_W^2} L_\mu J^\mu
\end{equation}
where $g_W$ is the $\mathit{SU(2)}$ coupling, $m_W$ the \w mass,
$L_\mu$ the leptonic current of the \tl, and $J_\mu$ a leptonic or
hadronic current dependent upon the decay. The \tl current $L_\mu$ is
$\bar{u}_\tn \gamma_\mu (1 - \gamma^5) u_\tau$ where $u_\tau$ is the
\tl spinor, dependent upon momentum and helicity, and $\bar{u}_\tn$ is
the \tl neutrino spinor. A full list of the available $J_\mu$ currents
modeled in \pythia is given in Table~\ref{tab:decay}.

For the two-body decays of the \tl, $J_\mu$ is $fq$, where $f$ is a
constant and $q$ is the momentum of the hadron. For the three-body
leptonic decays, $\tauto e^- \enb$ and $\tauto \mu^- \mnb$, $J_\mu$ is
of the same form as $L_\mu$. Two hadronic three-body decay models are
available, a decay via a vector resonance and a decay via a vector and
scalar resonance.

Four-body \tl decays in \pythia are implemented in four different
models. The primary four-body decays, $\tauto \pi^0 \pi^0 \pi^-$ and
$\tauto \pi^- \pi^- \pi^+$, are modeled using the CLEO fit. These
decays can also be performed using a generic hadronic four-body model
from Decker, et al. The four-body decays with kaons in the final state
are calculated using a model from Finkemeier and Mirkes.

The five-body decays of the \tl to pions are produced with the
Novosibirsk model, a phenomenological fit of four pion production from
electron-positron annihilation. The six-body decays of the \tl are
handled through a model proposed by K\"uhn and W\c{a}s.

\begin{figure}
  \begin{center}
    \ifdraft
    {\includesvg[width=\columnwidth,name=fig2,pretex=\footnotesize]{decay}}
    {\includegraphics[width=\columnwidth]{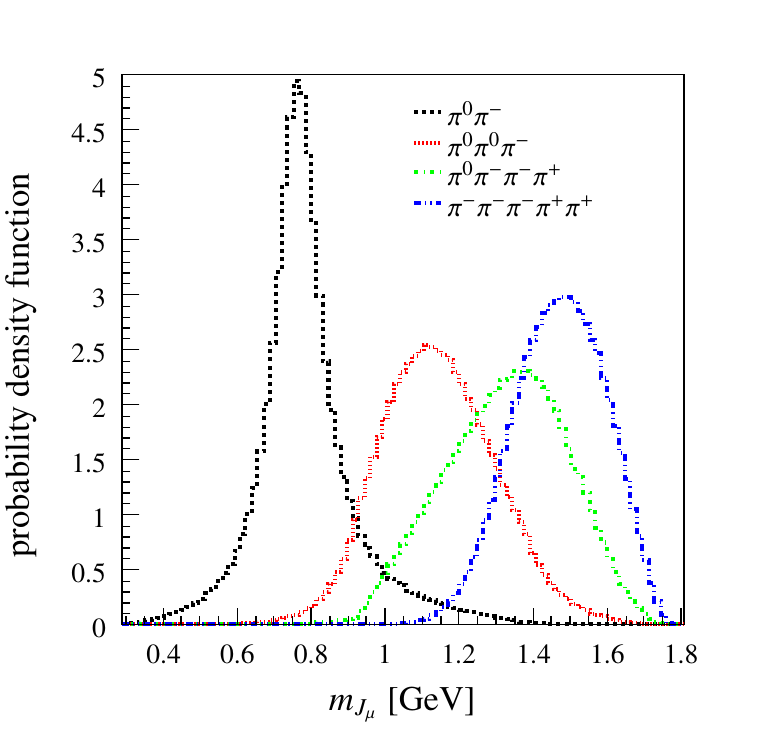}}
    \caption{Distribution of the invariant mass of the hadronic
      current for the dominant three-body $\pi^0 \pi^-$ (dashed
      black), four-body $\pi^0 \pi^0 \pi^-$ (dotted red), five-body
      $\pi^0 \pi^- \pi^- \pi^+$ (dash-dotted green), and six-body
      $\pi^- \pi^- \pi^- \pi^+ \pi^+$ (dash-dot-dotted blue) decays of
      the \tl.\label{fig:decay}}
  \end{center}
\end{figure}

A comparison of the hadronic invariant mass distributions, generated
with \pythia, between the dominant three, four, five, and six-body
decays of the \tl is given in Figure~\ref{fig:decay}. The $\tauto
\pi^0 \pi^0 \pi^-$ distribution was generated using the default CLEO
model. The expected upwards shift for higher multiplicity decays can
be seen, as well the distinct difference in shapes due to the
differing propagators of the models.

\begin{table*}
  \small
  \begin{center}
    \begin{tabular}{c|ll|c|l}
      \toprule
      \multicolumn{1}{c}{Multiplicity} &
      \multicolumn{2}{c}{Model} &
      \multicolumn{1}{c}{\me} &
      \multicolumn{1}{c}{Decay Products} \\
      \midrule[\heavyrulewidth]
      $2$-body                  & single hadron                           &                                     & 1521                  & $\pi^-$, $K^-$ \\
      \midrule                                                                                                                          
      \multirow{3}{*}{$3$-body} & leptonic                                &                                     & 1531                  & $e^- \enb$, $\mu^- \mnb$ \\
                                & K\"uhn and Santamaria                   & \cite{kuhn1}                        & 1532                  & $\pi^0 \pi^-$, $K^0 K^-$, $\eta K^-$ \\
                                & Finkemeier and Mirkes                   & \cite{finkemeier1}                  & 1533                  & $\pi^- \bar{K}^0$, $\pi^0 K^-$ \\
      \midrule                                                                                                  
      \multirow{6}{*}{$4$-body} & CLEO                                    & \cite{cleo}                         & 1541                  & $\pi^0 \pi^0 \pi^-$, $\pi^- \pi^- \pi^+$ \\
                                & \multirow{2}{*}{Finkemeier and Mirkes}  & \multirow{2}{*}{\cite{finkemeier2}} & \multirow{2}{*}{1542} & $K^- \pi^- K^+$, $K^0 \pi^- \bar{K}^0$, $K_S^0 \pi^- K_S^0$, $K_L^0 \pi^- K_L^0$, $K_S^0 \pi^- K_L^0$, \\
                                &                                         &                                     &                       & $K^- \pi^0 K^0$, $\pi^0 \pi^0 K^-$, $K^- \pi^- \pi^+$, $\pi^- \bar{K}^0 \pi^0$ \\
                                & \multirow{2}{*}{Decker, et al.}         & \multirow{2}{*}{\cite{decker}}      & \multirow{2}{*}{1543} & $\pi^0 \pi^0 \pi^+$, $\pi^- \pi^- \pi^+$, $K^- \pi^- K^+$, $K^0 \pi^- \bar{K}^0$, $K^- \pi^0 K^0$, \\
                                &                                         &                                     &                       & $\pi^0 \pi^0 K^-$, $K^- \pi^- \pi^+$, $\pi^- \bar{K}^0 \pi^0$, $\pi^- \pi^0 \eta$ \\
                                & Jadach, et al.                          & \cite{jadach}                       & 1544                  & $\gamma \pi^0 \pi^-$ \\
      \midrule                                                                                                  
      $5$-body                  & Novosibirsk                             & \cite{novosibirsk}                  & 1551                  & $\pi^0 \pi^- \pi^- \pi^+$, $\pi^0 \pi^0 \pi^0 \pi^-$ \\
      \midrule                                                                                                  
      $6$-body                  & K\"uhn and W\c{a}s                      & \cite{kuhn2}                        & 1561                  & $\pi^0 \pi^0 \pi^- \pi^- \pi^+$, $\pi^0 \pi^0 \pi^0 \pi^0 \pi^-$, $\pi^- \pi^- \pi^- \pi^+ \pi^+$ \\
      \bottomrule
    \end{tabular}
    \caption{Summary of available \tl decay models in \pythia $8$
      sorted by multiplicity. For each model the internal \pythia
      matrix element mode identifier (\me) is given, as well as the
      decays available through the model and the reference of the
      model. The implicit \tn is omitted for
      brevity.\label{tab:decay}}
  \end{center}
\end{table*}

\section{Interface}

Documentation for \tl decays in \pythia can be found under the {\it
  Tau Decays} subsection of {\it Particle Decays} in the
\href{http://home.thep.lu.se/~torbjorn/pythia81html/Welcome.html}{\pythia
  HTML manual}. The \tl decay mechanism is configured by setting the
mode {\tt sophisticatedTau} of {\tt ParticleDecays}. This mode default
is {\tt 1}, where the decays are performed using the sophisticated \tl
decay machinery. When sophisticated decays are not necessary, or a
comparison with the isotropic \pythia model is needed, the mode can be
set to {\tt 0}.

If the mode is set to {\tt 1}, and a LHEF file is read, the spin of the
\tl will be set to the SPINUP information. However, if no SPINUP
information is provided, and the \tl is from a known production
process, \pythia will automatically calculate the polarization. If the
\tl is from an unknown process the sophisticated decay will be
performed assuming the \tl is unpolarized.

It is possible to force the polarization of the \tls and override
either the \pythia calculated polarization or the polarization from a
LHEF file. If the mode is set as {\tt 2}, all \tls produced from a
parent with an identification code specified by {\tt tauMother} will
be given a polarization set by the parameter {\tt tauPolarization}. If
mode {\tt 3} is selected, all \tls will be produced with the
polarization set by {\tt tauPolarization}.

The models used to decay the \tls for each channel can be changed by
switching the matrix element mode of the channel. The syntax takes the
form {\tt 15:channel:meMode = mode} where {\tt 15} is the \tl particle
identification code, {\tt channel} specifies the decay channel number
as listed under {\it Particle Data}, and {\tt mode} is the new matrix
element mode. For example, the default CLEO model used for the $\tauto
\pi^0 \pi^0 \pi^-$ decay can switched to the Decker, et al.\ model
using {\tt 15:9:meMode = 1543}. However, caution is advised in
switching models, as it is possible to set a channel to use an
incompatible matrix element, causing unpredicted behavior.

\section{Implementation}

The implementation of the \tl decay machinery in \pythia is intended
to be complete and fast, yet easily extensible for new decay models or
production processes. A helicity framework is available in the \pythia
source files {\tt HelicityBasics} where methods for calculating the
helicity dependent wavefunctions for particles, both fermions and
bosons, is provided, as well as the corresponding $\gamma$-matrices.

Using this framework, the helicity matrix element for the decay of a
\tl into a pion,
\begin{equation}
  \me = \bar{u}_\tn \gamma_\mu (1 - \gamma^5) u_\tau f_\pi q_\pi^\mu
\end{equation}
can be written in pseudo-code as,
\begin{equation}
  \begin{aligned}
    \me = \sum_\mu &\tn\mathtt{.waveBar(} \lambda_\tn
    \mathtt{)*GammaMatrix(} \mu \mathtt{)} \\[-0.4cm]
    & \mathtt{*(1 - GammaMatrix(5))*}
    \tau\mathtt{.wave(} \lambda_\tau \mathtt{)*} f_\pi \\
    & \mathtt{*GammaMatrix(4)(} \mu
    \mathtt{,} \mu \mathtt{)*Wave4(}q_\pi\mathtt{)(} \mu \mathtt{)} \\
  \end{aligned}
\end{equation}
where $\gamma^4$ provides the $+---$ metric.

Both the production matrix elements and \tl decay matrix elements are
written using this framework and are provided in the {\tt
  HelicityMatrixElements} files of the \pythia source. For the \tl
decay matrix elements, it is only necessary to implement the $J_\mu$
current. The spin correlation and decays of the \tls are implemented
in the {\tt TauDecays} files of the \pythia source.

\section{Conclusion}

Fully modeled \tl decays with spin correlations are now available in
\pythia, using fully extensible code. Further development of \tl
production mechanisms and decay models is underway, and feedback on
requested features is welcome.

\section{Acknowledgments}

The author would like to acknowledge Torbj\"orn Sj\"ostrand and the
\pythia team for all of their invaluable help. The funding for this
project was provided by Lund University and MCNet through a Marie
Curie grant, contract MRTN-CT-2006-035606.

\nocite{*}
\bibliographystyle{elsarticle-num}
\bibliography{proceedings}
\end{document}